# Preliminary Design Study of the Hollow Electron Lens for LHC

D. Perini, C. Zanoni

**Summary**

A Hollow Electron Lens (HEL) has been proposed in order to improve performance of halo control and collimation in the Large Hadron Collider in view of its High Luminosity upgrade (HL-LHC). The concept is based on a beam of electrons that travels around the protons for a few meters. The electron beam is produced by a cathode and then guided by a strong magnetic field generated by a set of superconducting solenoids. The first step of the design is the definition of the magnetic fields that drive the electron trajectories.
The estimation of such trajectories by means of a dedicated MATLAB® tool is presented. The influence of the main geometrical and electrical parameters are analysed and discussed. Then, the main mechanical design choices for the solenoids, cryostats gun and collector are described. The aim of this paper is to provide an overview of the preliminary design of the Electron Lens for LHC. The methods used in this study also serve as examples for future mechanical and integration designs of similar devices.

## 1 Introduction

Hollow electron collimation is a novel technique for beam collimation and halo scraping [1, 2] based on a cloud of electrons traveling around the proton axis for a few meters. The electrons are emitted by a cathode and then compressed and confined into a long strong solenoid. The electrons finally dissipate their energy on a metallic collector with an active cooling system.
Such a system has recently been proposed for the Large Hadron Collider at CERN [3]. An overview of this kind of devices and their technological status is presented in [4].
The electron cloud is hollow and axisymmetric and overlaps with the proton beam halo. The particles of the halo are kicked transversally, while the core remains unaltered thanks to the fact that a hollow axially symmetric charge distribution does not generate any electromagnetic field in the inside empty space. There are several advantages in employing an electron cloud around the beam, [5]. The most evident one is that the electron beam can be placed at any distance from the proton beam centre, without any limitation on the minimal distance.
The trajectory of the electrons is driven by the magnetic field generated by a set of solenoids. The solenoid in which the cathode is located is called *gun solenoid,* the long one that confines the e-cloud is called *main solenoid*. Intermediate solenoids are also foreseen in order to keep the electrons on the defined path, Figure 1.
There will be two of these devices, one per beam, and they will be installed in proximity of LHC point 4. In this location the LHC beam-beam distance is 420 mm. This dimension calls for a compact design that fits in the small available space. Another basic characteristic is the robustness. A solenoid quenching often would spoil the availability of the LHC machine. For this reason the solenoids are largely oversized in terms of nominal current with respect to the critical one.

The field level determines the radius of the e-cloud. The dimensions of the electron beam in two points along its path follow the equation:

$$\frac{r_0}{r_1} = \sqrt{\frac{B_1}{B_0}} \qquad 1)$$

Where $r_0$ and $r_1$ are the radii of the electron beam in point 0 and 1 and $B_0$ and $B_1$ are the magnetic field in points 0 and 1 respectively.

The gun solenoid is divided in two parts: one tuneable field solenoid around the e-gun cathode and a constant field solenoid in line with the first one. The dimensions of the electron beam in the main solenoid can be tuned varying the level of the field in the tuneable solenoid following the equation 1). The constant field solenoid together with the main solenoid and the intermediate bending solenoids define the trajectory of the electrons once they have left the emission area. In other words, in order to tune the electron beam size one needs to change the field level of the tuneable solenoid alone while all the other coils stay at their nominal current level.

The table below gives the nominal dimensions of the electron beam in the main solenoid and the dimensions of the emitting cathode in the electron gun.

*Table 1: Dimensions of the hollow electron beam and of the emitting cathode. As an example we give the dimensions that can be obtained changing the e-gun tuneable solenoid from the nominal level of 0.2 T to 0.1 T.*

| | |
|---|---|
| Nominal magnetic field of the main solenoid | 4 T |
| Inner radius of the hollow electron beam @ nominal fields | 0.9 mm (3 σ) |
| Outer radius of the hollow electron beam @ nominal fields | 1.8 mm (6 σ) |
| Nominal magnetic field in the e-gun cathode | 0.2 T |
| Inner diameter of the cathode | 8.05 mm |
| Outer diameter of the cathode | 16.10 mm |
| Inner radius of the hollow electron beam @ 4 T with 0.1 T @ cathode | 0.635 mm |
| Outer radius of the hollow electron beam @ 4 T with 0.1 T @ cathode | 1.275 mm |
| Nominal current at the cathode | 5 A |

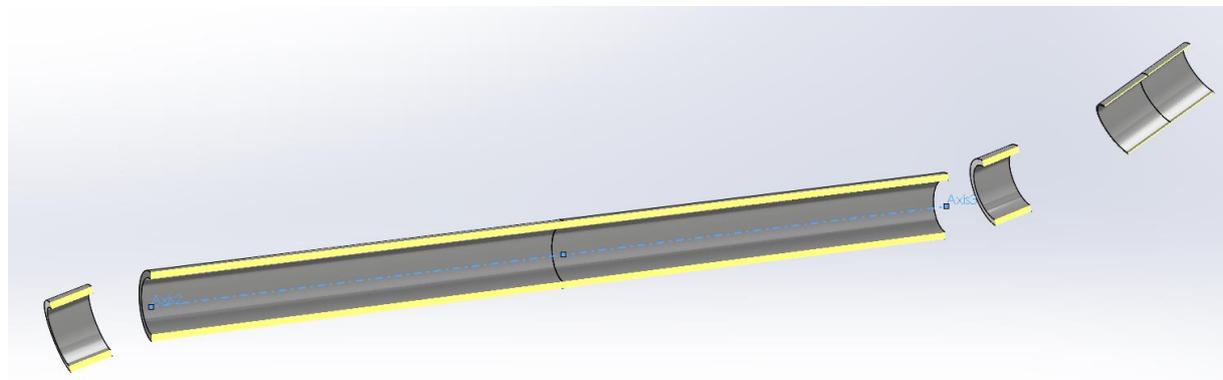

*Figure 1: Open view of the solenoids configuration*

## 2  Configuration of Solenoids Driving the Electron Cloud

This section describes the assumptions and calculations made to determine the trajectory of the electrons and define the optimal configuration of the solenoids. We evaluated as well the sensitivity of the system with respect to the main geometrical parameters in order to estimate the construction mechanical tolerances and define a system of a correction coils.

The numerical computations are carried out with the code MATLAB® and occasionally with Comsol 5.2®. MATLAB® is chosen as it allows the maximum flexibility in exploring different types of design.

The electrons are emitted by the cathode and travel toward the main solenoid. In the simulations an emission point is located on the e-gun solenoid axis and we check that the trajectory of the electrons overlaps the axis of the main solenoid. The electrons follow the flux lines of the fields generated by the set of solenoids. As a general rule from the experience with similar applications, the magnetic field experienced by the electrons should always be $\geq 0.1$ T otherwise the trajectory is not stable.

The following assumptions are introduced to simplify the computations:

  a. We consider a single particle and not the cloud of electrons. The mutual interaction between particles is also neglected as well as the influence of the proton beam and of the metal vacuum pipe.
  b. the relativistic correction is not applied;
  c. ambient disturbances are neglected.

The goal of this part of the work is to provide easily, quickly and in a flexible way guidelines for mechanical design. The results do not substitute a full assessment of beam dynamics. We consider this as the starting point to define the structures. It is expected that the real electron beam in the uniform field volumes will follow almost the same path and the difference will be small compared to the mechanical tolerances. As an order of magnitude, all the tolerances piled up can give a total error in the position of the magnets of the order of a few millimetres.

Several options for the design have been explored. The ideal system would be an S-shaped solenoid that goes from the cathode to the 3 m long straight section around the proton beam and then to the collector. Indeed, this is not an option since the solenoids have to be straight and the space for the proton beam pipe has to be guaranteed.

The design concept analysed in the following foresees the presence of two short intermediate solenoids at the inlet and outlet of the main solenoid. These solenoids are tilted and are powered with the same current of the main one in order to bend the electrons with a strong field.

The gun solenoid is divided in two parts. One part with a higher field (~0.4 T) creates together with the intermediate and the main solenoid the correct field path. The second part surrounds the cathode and is at lower field (0.2 T) to guarantee the correct compression factor.

In the rest of this document, the solenoid names will be shortened as follows:

  - MS: main solenoid
  - GSi: gun solenoid with higher field
  - GSc: gun solenoid hosting the cathode
  - BSg: intermediate bending solenoid, gun side
  - BSc: intermediate bending solenoid, collector side

## 2.1 Coarse Estimation of the Electron Trajectory and its limits

The configuration of the solenoids is derived by the previous examples of electron lenses in accelerators [6,7,8] but is changed and optimized for the HL-LHC machine and its tight requirements due to the high energies involved.
The coordinate system used for the analysis is as follows:

- o the origin is in the centre of the main solenoid;
- o the z axis is parallel to the main solenoid symmetry axis;
- o the y axis, together with the z, defines the plane in which lay both the gun and the main solenoid axes;
- o the x axis completes the set according to the right-hand rule.

The motion equation of a charged particle in a magnetic field is:

$$m \ddot{\boldsymbol{x}} = q \dot{\boldsymbol{x}} \times \boldsymbol{B}(\boldsymbol{x}, t) \qquad 2)$$

where $\boldsymbol{x}$ is the position, m the mass, q the charge and $\boldsymbol{B}(\boldsymbol{x}, t)$ the magnetic field. This field is derived integrating the Biot-Savart law for each solenoid, which is usually expressed as:

$$\boldsymbol{B}(\boldsymbol{x}, t) = \frac{\mu_0}{4\pi} \iiint \frac{(j \, \hat{\boldsymbol{\iota}} \, dV) \times \boldsymbol{r}}{|\boldsymbol{r}|^3} \qquad 3)$$

Where $\mu_0$ is the vacuum permeability, $j$ is the current density, $\hat{\boldsymbol{\iota}}$ is the versor of the solenoid wire element, $V$ is the solenoid volume and $\boldsymbol{r} = \boldsymbol{x} - \boldsymbol{l}$ is the vector from the position in which the field is evaluated to the wire element.
In this analysis the integral is:

$$\boldsymbol{B}(\boldsymbol{x}, t) = \frac{\mu_0}{4\pi} j \, \Delta R_s R_s \int_{L/2}^{-L/2} \int_{-\pi}^{\pi} \frac{d\theta dh \hat{\boldsymbol{\iota}} \times \boldsymbol{r}}{|\boldsymbol{r}|^3} \qquad 4)$$

Where $\Delta R_s$ is the difference between the internal and external solenoid radius, $R_s$ is the nominal radius and L is the solenoid length.
Using a double integral instead of a triple means that solenoid thickness is considered small. $R_s$ is then assumed equal to the average radius. The effect of this assumption is shown in Figure 3, where the magnetic field has been estimated in some areas with both a double and a triple integral. The field difference is below 1 % in the analysed area, the one in which lay the electron trajectory. The computations with the triple integral instead of the double one take approximately 10 – 15 times longer.

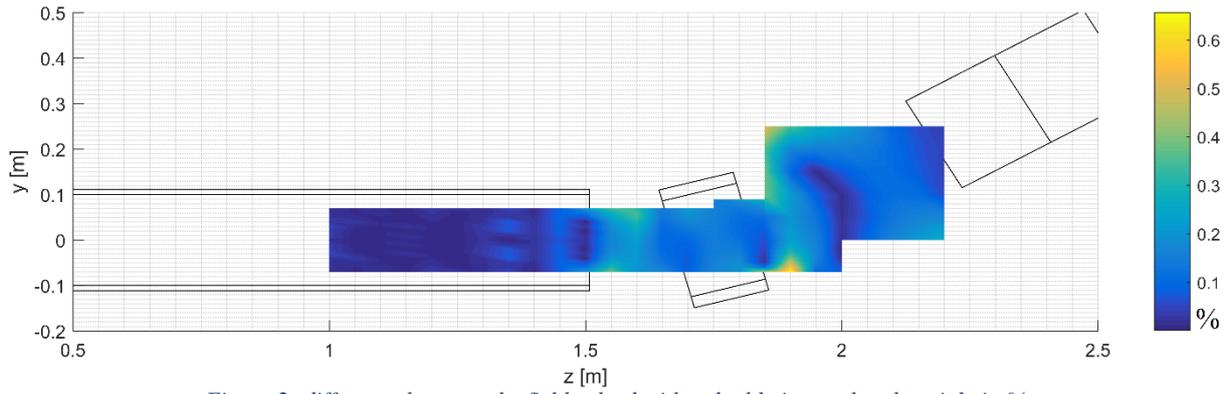
*Figure 2: difference between the field solved with a double integral and a triple in %*

When a solenoid is not aligned with one of the axes of the coordinate system, a proper roto-translation is applied.
The magnetic field is also estimated with Comsol® that uses a finite element approach. Figure 3 shows the field lines obtained with this code. Figure 4 gives the difference in field magnitude between the integrated Biot-Savart law and Comsol®. The difference close to the axis of the solenoids, the region of the electron path, is limited to less than a few mT (< 1 %).

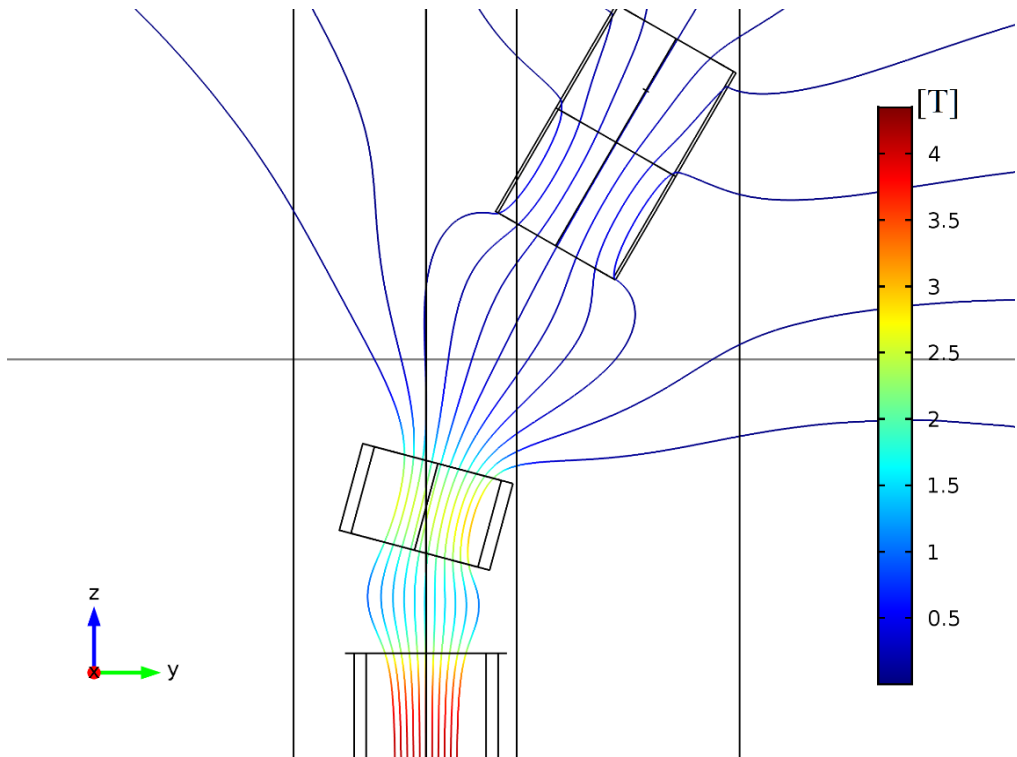
*Figure 3: field lines as estimated in Comsol*

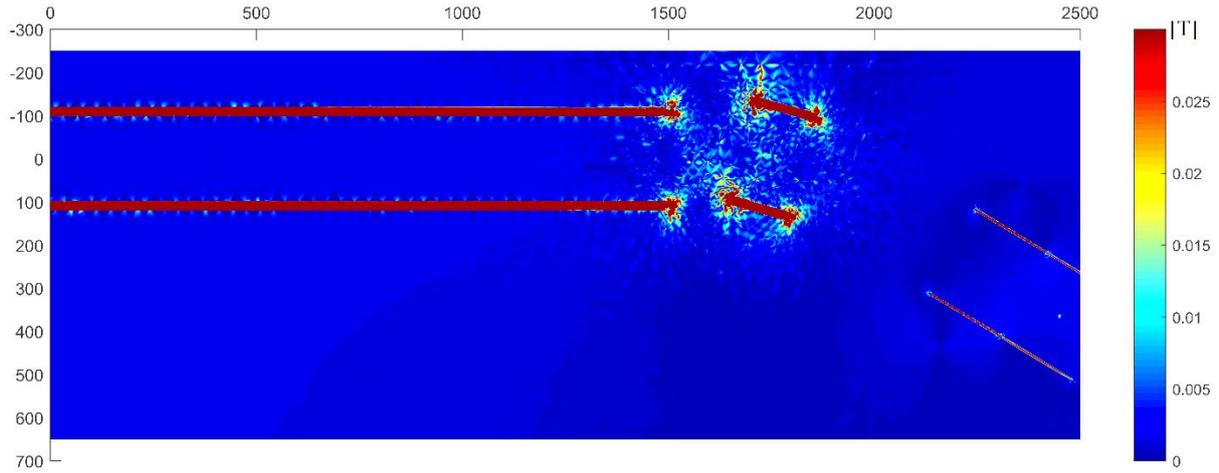
*Figure 4: difference in field magnitude between Comsol® and the Biot-Savart law integrated in MATLAB®*

There are two possibilities to integrate the particle motion equation. The first option is the numerical solution of the full Newton-Lorentz set of equations through the MATLAB® built-in functions such as ode23s and ode23t. The possibility of integrating the equations according to the guiding centre approximation [9] has also been explored.

The second option is the implementation of a solution algorithm optimized for charged particle motion, such as the Boris method [10, 11]. The equations of the Boris scheme, without electric field, are:

$$\boldsymbol{x}_{k+1} = \boldsymbol{x}_k + \Delta t\, \boldsymbol{v}_{k+1/2}$$
$$\boldsymbol{v}_{k+1/2} = \boldsymbol{u}'$$
$$\boldsymbol{u}' = \boldsymbol{u} + \big(\boldsymbol{u} + (\boldsymbol{u} \times \boldsymbol{h})\big) \times \boldsymbol{s}$$
$$\boldsymbol{u} = \boldsymbol{v}_{k-1/2}$$
$$\boldsymbol{h} = \Delta t \frac{q}{2\,m} \boldsymbol{B}_k$$
$$\boldsymbol{s} = \frac{2\,\boldsymbol{h}}{(1+h^2)}$$

5)

Where $\boldsymbol{x}_k$ is the position vector at the k-step, $\boldsymbol{v}_{k+1/2}$ is the velocity vector between step k and k+1, $\boldsymbol{B}_k$ is the magnetic field in $\boldsymbol{x}_k$, q is the electron charge, m its mass and $\Delta t$ is the time step. It has been noted that most of the computational time is spent in calculating the magnetic field at each step. In order to speed this up, the magnetic field is updated only once every $n$ time steps where $n$ depends on the maximum change of magnetic field components:

| $\max(B_{x/y/z,k} - B_{x/y/z,k-n})/n$ | $n$ |
|---|---|
| $> 1.5\ 10^{-4}\ T$ | 5 |
| $5\ 10^{-5} \div 1.5\ 10^{-4}\ T$ | 12 |
| $0 \div 5\ 10^{-5}\ T$ | 40 |

The Boris method results being both the fastest and the most accurate approach, as also suggested by the literature [11]. Preliminary analysis also showed that this is the only approach in which the kinetic energy stay constant, as should happen in absence of electric fields.

The guiding centre approximation is very promising but suffers of the same problems as the direct integration: the energy is dissipated. Furthermore, the magnetic field changes quickly at the entry section of the main solenoid and this can create further problems for the accuracy of the solution.

The initial particle velocity used for simulation is $5.935 \times 10^7$ m/s, roughly equivalent to 10 keV (without relativistic correction), along the gun axis. The initial position is the centre of the GSc, see cross in Figure 5.

## 2.2 Optimal Configuration and Sensitivity

An optimal configuration of the solenoids will send the electrons inside the MS over the axis of the solenoid. The distance between the electron beam and the solenoid axes, if different from zero, can be represented by the two components x and y. As first approximation these two components are considered independent. The x component of the distance may seem surprising in an axis-symmetrical system. It is the effect of the motion equations, which have a cross product between velocity and field. A discussion of a similar application can be found in [8].

A linear trajectory is not necessarily perfectly parallel to the MS axis. The angles of the inclination are estimated, but not optimized. In quantitative terms, the objective function to be zeroed is the average distance of the calculated trajectory in the range of z [-2 m, +2 m].

In principle, all the geometrical parameters can be adjusted. This would then determine a problem with a high number of degrees of freedom. We therefore decided to fix the radiuses and the nominal currents according to considerations of other nature such as field compression factor, magnetic energy and reasonable space available.

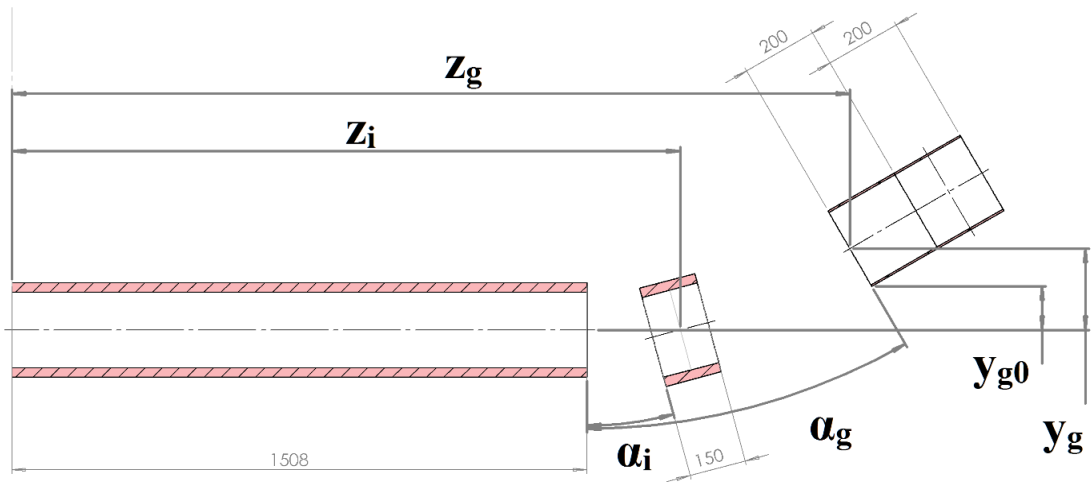

*Figure 5: solenoids free parameters*

The residual free geometrical parameters are shown in Figure 5. The other dimensions are detailed in the second part of this document. The key parameters for the trajectory are:

| Solenoid | Requirement | Other |
|---|---|---|
| MS | 4 T | 159.5 A |
| BS | 159.5 A | ~2.5 T |
| GSi | 94 A | ~0.4 T (centre) |
| GSc | 0.2 T (centre) | 25.9 A |

At this point, in order to prevent excessive curvature of the electrons trajectory, the gun is kept as close as possible to the proton beam pipe. The BSg and BSc angle $\alpha_i$ is the highest allowed by beam pipe integration. Also the z position of these solenoids is minimized in order to keep the system as compact as possible. The angle $\alpha_g$ is fixed, but the combination of this value with $z_i$ is discussed in the following paragraphs.

In summary, the optimized parameters are:

| | |
|---|---|
| $y_{g,0}$ | 0.115 m |
| $\alpha_g$ | $\pi/6$ |
| $\alpha_i$ | $\pi/12$ |
| $z_i$ | 1.753 m |

From which the distance of the GS is:

$$y_g = y_{g,0} + R_{gun}\cos\alpha_g \qquad 6)$$

where $R_{gun}$ is the external radius of the gun solenoid.

The correlation between distance and angle of the GS is shown in Figure 6. The steeper is the entrance of the electrons into the BSg, the closer the GS can be placed to the MS. The advantage of having a more compact system is paid in terms of stability of the electrons trajectory, which is more bent.

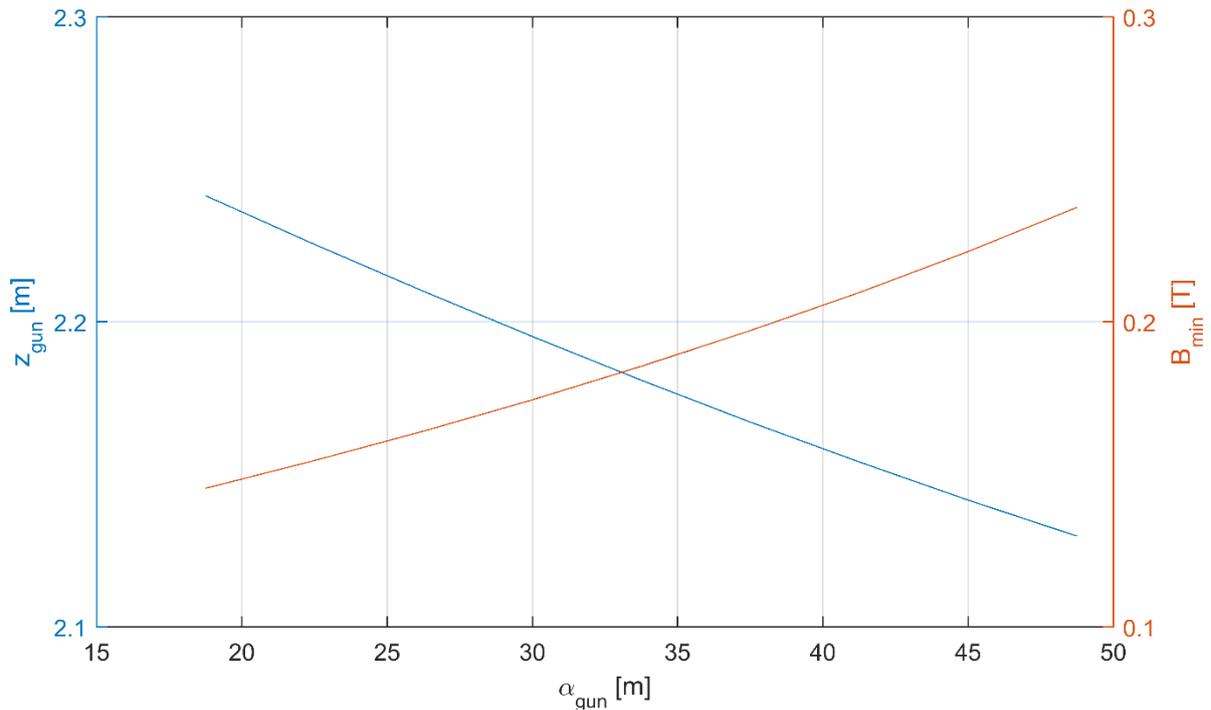

*Figure 6: correlation between distance of the GS and their angle (blue). The minimum field encountered with that configuration is also estimated (red).*

Figure 10 shows the curvature of the trajectory in the plane z-y, where the high order oscillations have been filtered. The angle $\alpha_g = \pi/6$ is chosen to be as small as possible, to minimize the trajectory bending, and to have the curvature of the same sign between the GSi and the BSg. It is worth noting that the peak of curvature is independent from the gun inclination (in first approximation, it depends only on the bending part). Finally the value of $z_g$ is fixed to 2.190 m.

With this set of parameters, the trajectory is the one shown in Figure 7, where the top and bottom lines represent the external edge of the electron cloud calculated according to the compression factor formula. Figure 8 is zoomed on the bending part of the trajectory.

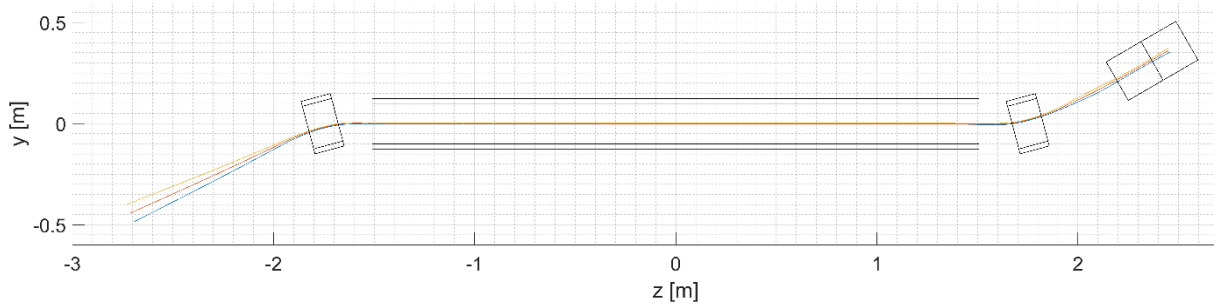

Figure 7: trajectory estimated with the optimal parameters. Top and bottom lines represent the external edge of the electron cloud calculated according to the compression factor formula.

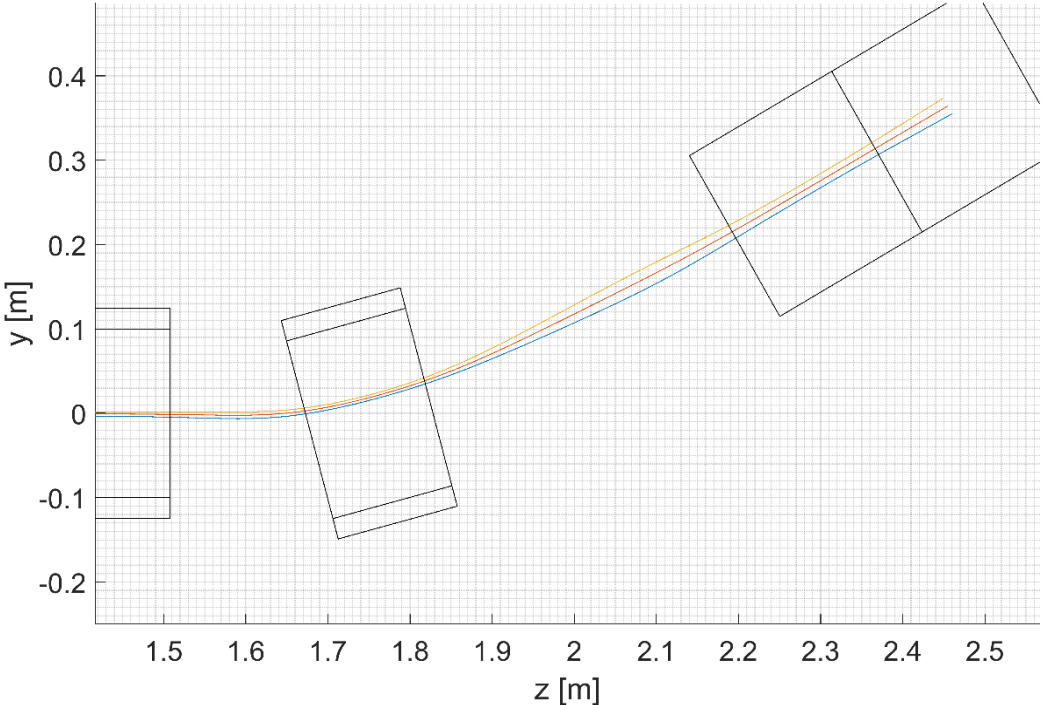

Figure 8: trajectory estimated with the optimal parameters, zoom on the inlet bending

Figure 9 shows the magnetic field magnitude as measured on the simulated trajectory. Note that the field at the initial position requires the adjustment of the current of the GSc to obtain the required 0.2 T.

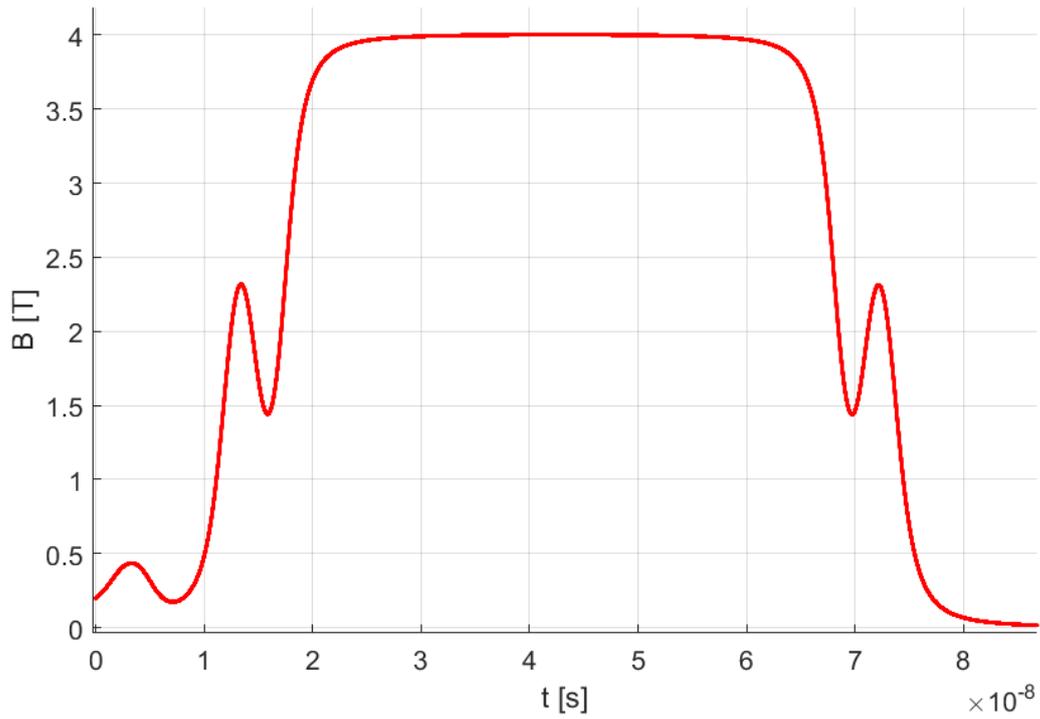
*Figure 9: magnetic field seen on the simulated central trajectory*

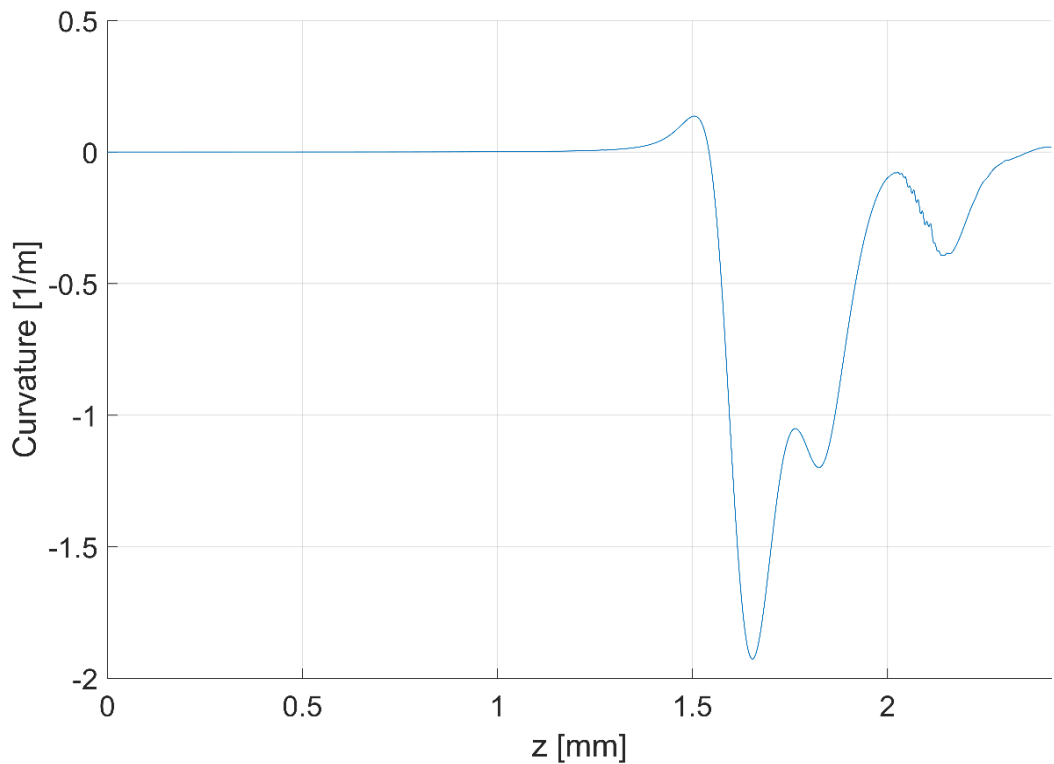
*Figure 10: curvature of the electron trajectory for $\alpha_g = \pi/6$*

Figure 11 shows the correlation between the position of the BS and the GS, important for the compactness of the structure. In facts, every translation of the BS determines an equal or greater translation of the GS.

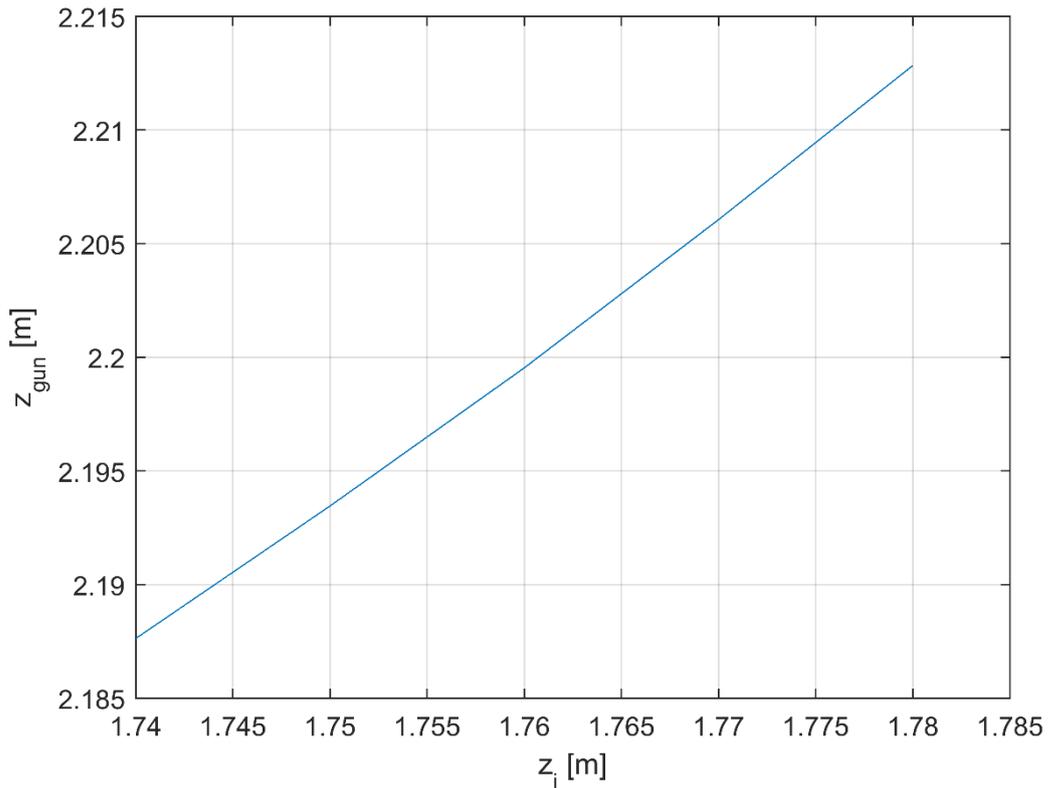

*Figure 11: correlation between the position of the BS and the GS*

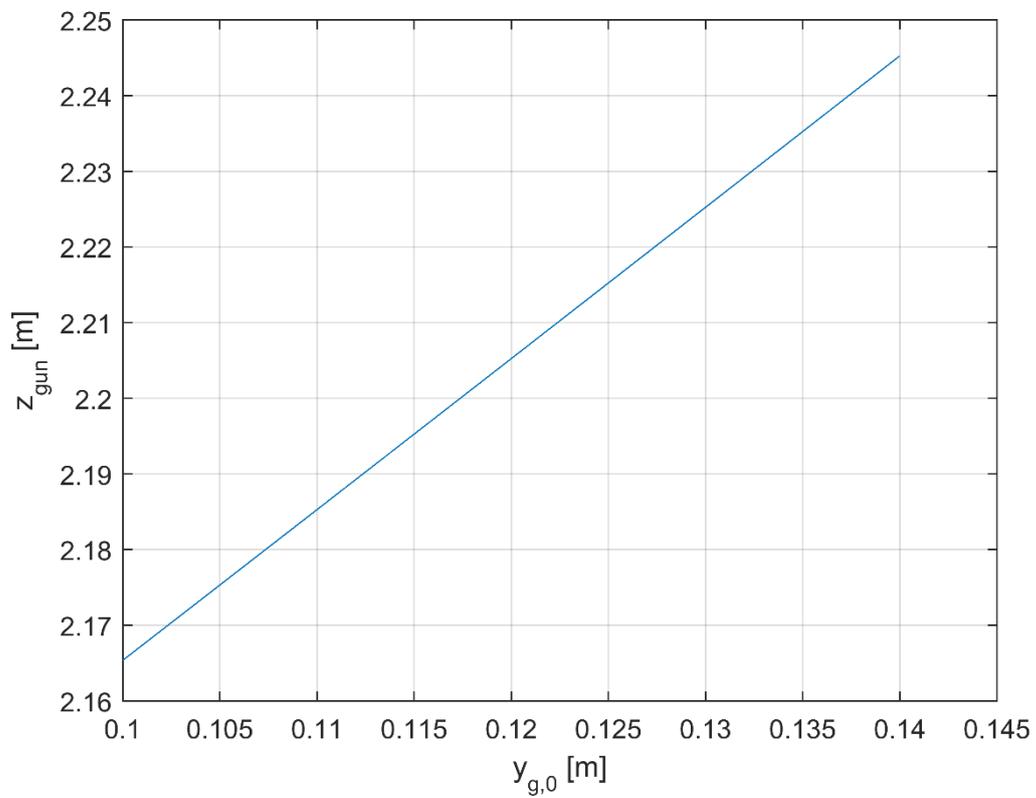

*Figure 12: correlation between the position of the GS along y and along z*

Figure 12 shows the correlation between the position of the GS along y and along z, which has consequences on the total length of the hollow electron lenses. Increasing the distance from the GS to the proton beam, means a longer structure in z, if the angle $\alpha_g$ is kept constant.

Table 2 gives the sensitivity of the trajectory position (i.e. of the linear part inside the MS) with respect to several input parameters. $x_{ct}$ and $y_{ct}$ are the mean distances of the trajectory from the solenoid axis, $\alpha_{ct}$ is the angle in the y-z plane.

*Table 2: sensitivity of trajectory position as a function of several design parameters*

|  |  | $x_{ct}$ | $y_{ct}$ | $\alpha_{ct}$ |
|---|---|---|---|---|
|  | *Units* | *1/m* | *1/m* | *1/rad* |
| z gun | m | 0.0017 | -0.0643 | -3.7e-6 |
| y gun | m | -0.0026 | 0.1247 | 2.8e-5 |
| x gun | m | 1 | 1.5e-5 | -1.8e-7 |
| $\alpha_g$ | rad | 0.0019 | -0.014 | -1.3e-5 |
| Current MS | A | -1.2e-6 | 6.0-6 | -4.8e-8 |
| Current gun (0.4 T) | A | -2.6e-7 | -9.3e-6 | -4.3e-8 |
| Current gun (0.2 T) | A | -1.2e-7 | -3.4e-6 | 3.3e-9 |
| z BSg | m | -0.0015 | 0.0393 | 0.00032 |
| angle BSg | rad | -4.70e-5 | -0.1039 | -0.0004 |

## 2.3 Electron Collector

At the end of their trajectory parallel to the LHC beam the electrons are bent off-axis by the bending solenoid. Then they continue their path and are stopped by the collector and deposit there their energy. The theoretical maximum power dissipated in the collector 50 kW. This value can be reduced by changing the potential of the collector [4], however, for the preliminary design we considered the total maximum power. A cylinder of magnetic steel that absorbs and opens the magnetic field lines surrounds the collector. The proposed HEL geometry do not foresee any solenoid around the collector. The electrons travel along the field flux lines and they are spread on a large surface before reaching the collector. The power of the electron beam is therefore distributed on a large area and the local temperature can stays within acceptable limits. A description of a similar design can be found in [12].

A set of numerical simulations have been run in Comsol® to assess:
- Functionality of collector without guiding solenoid.
- Preliminary position and shape of the collector.
- Size of the area on which the electrons deposit their energy.

The preliminary configuration for the collector is shown in Figure 13. The longitudinal dimensions have their origin in the centre of the main solenoid.

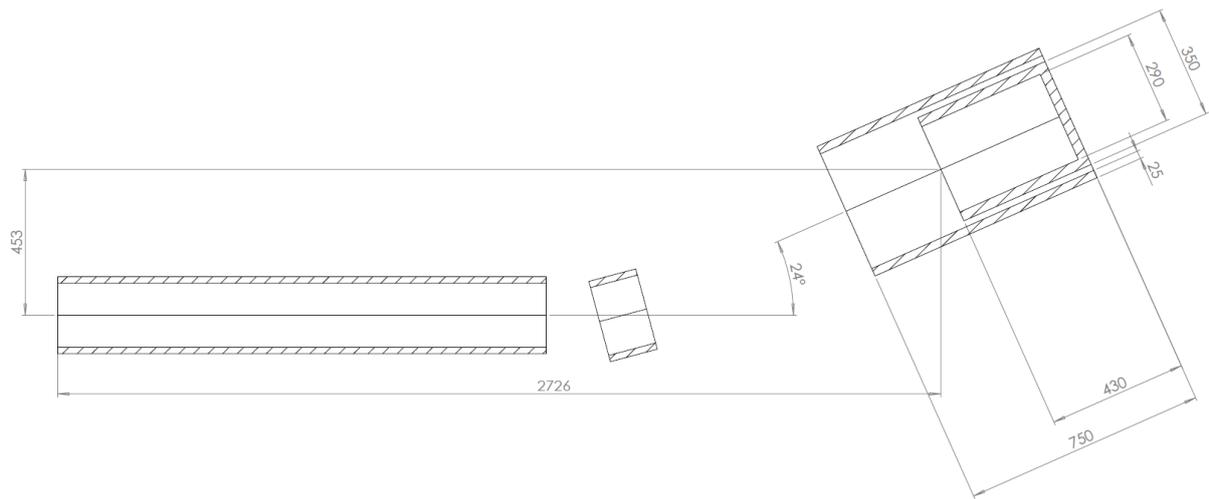

*Figure 13: preliminary configuration of the collector region*

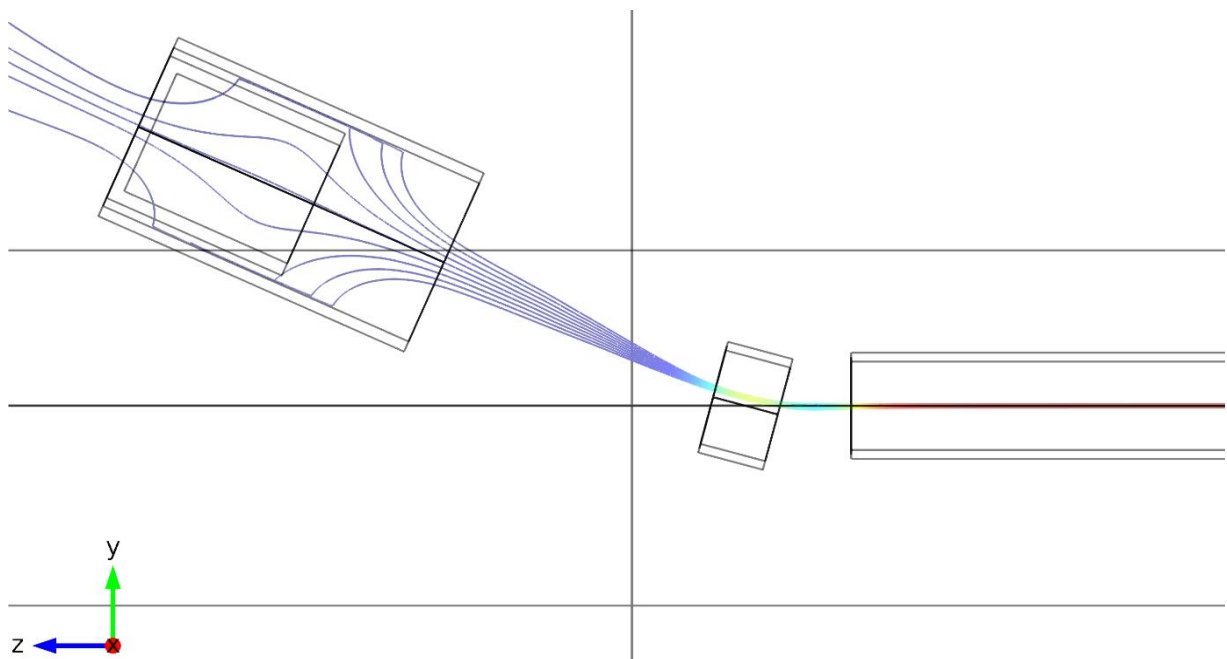

*Figure 14: field lines in the collector region*

Figure 14 shows the magnetic field generated with this configuration. In our simulation, the magnetic shield around the collector has been modelled with the ferrite non-linear material of the Comsol® database. We used as well a linear material with relative permeability 2500 and we obtained very similar results.

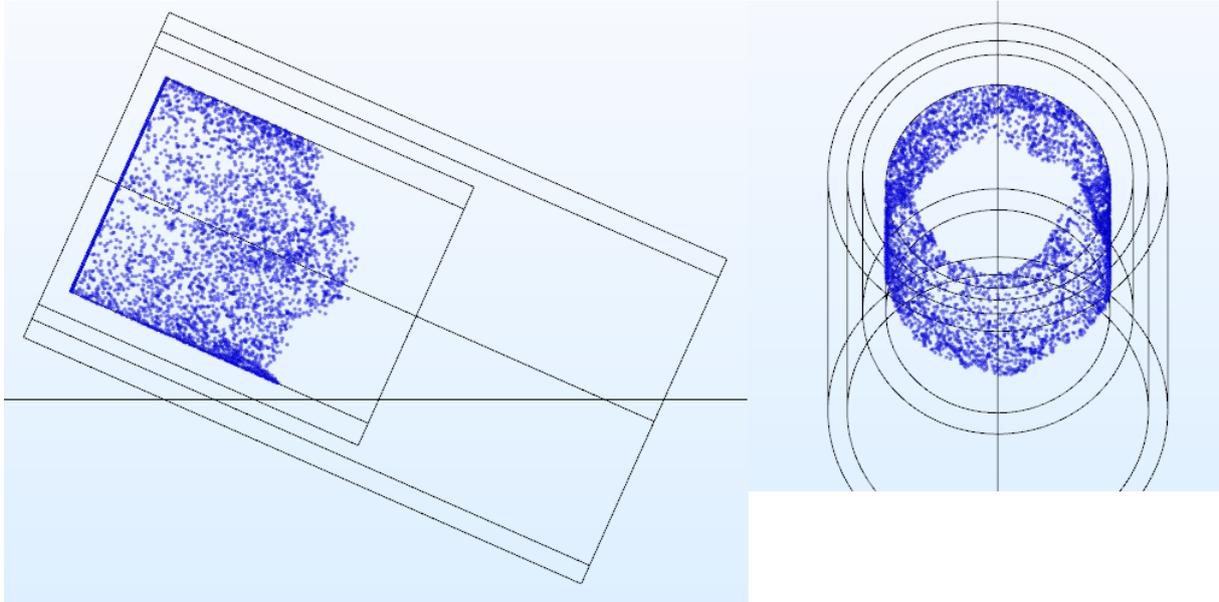

*Figure 15: position in which the electrons reach the collector (simulated with 5000 particles, equivalent to the 5 A current)*

Figure 14 shows the location in which the electrons deposit their energy (simulated with 5000 particles equivalent in mass and charge to 5 A current). Approximately 3/4 of the collector length directly absorb energy.

The configuration of the collector area can be further optimized, however the preliminary results provide useful data for its preliminary design. It is also recommended to foresee a mechanism that allows the adjustment of the ferrite shield position to optimize the energy deposition.

## 3 Preliminary Mechanical Design

### 3.1 General description

Once defined the optimal magnetic field configuration we can size the magnets and the services. This section describes the technological choices proposed for the HEL.
The main solenoid and the bending solenoids are superconducting. They are assembled inside cryostats cooled by liquid helium at 4 - 5 K and gas helium available in the LHC tunnel.
The e-gun solenoids work at a field level that could be obtained using normal resistive coils. A resistive solenoid of 0.4 – 0.5 T foresees a huge copper winding dissipating 10-15 kW power and working at 400 – 500 A. Given the fact that the liquid helium is available in the tunnel and already used in the rest of the HEL coils, we think that a superconducting solenoid is the best choice for the e-gun magnet.as well. In the long run it will pay back the initial investment.
The HEL hardware is therefore made of a set of superconducting coils assembled in cryostats. The cryostats have a central bore to house the room temperature vacuum chamber where the LHC beam and the hollow electron beam travel.

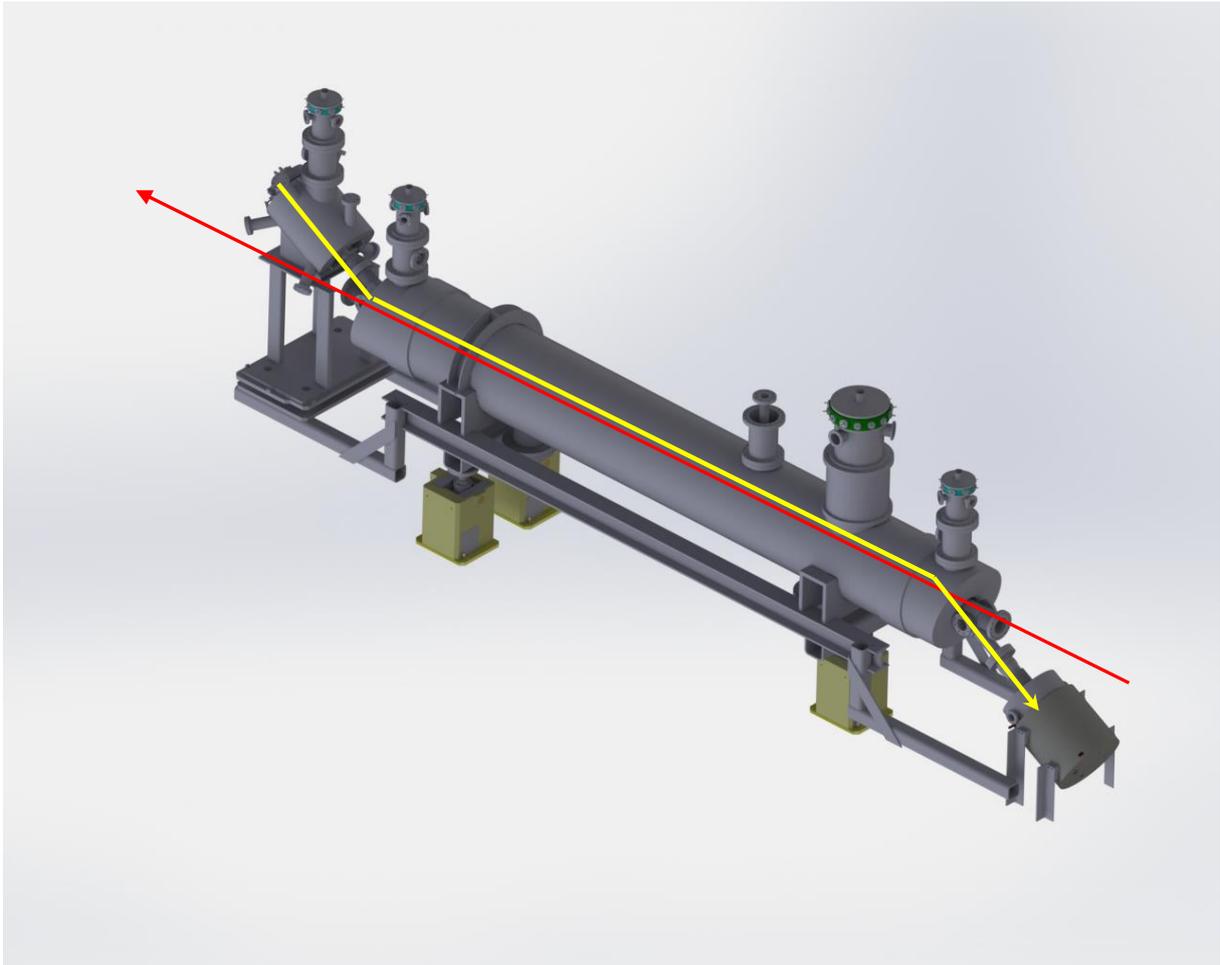

*Figure 16: 3D view of the HEL system. The red arrow indicates the travelling path of the LHC beam, the yellow arrow the path of the electrons from the e-gun to the collector.*

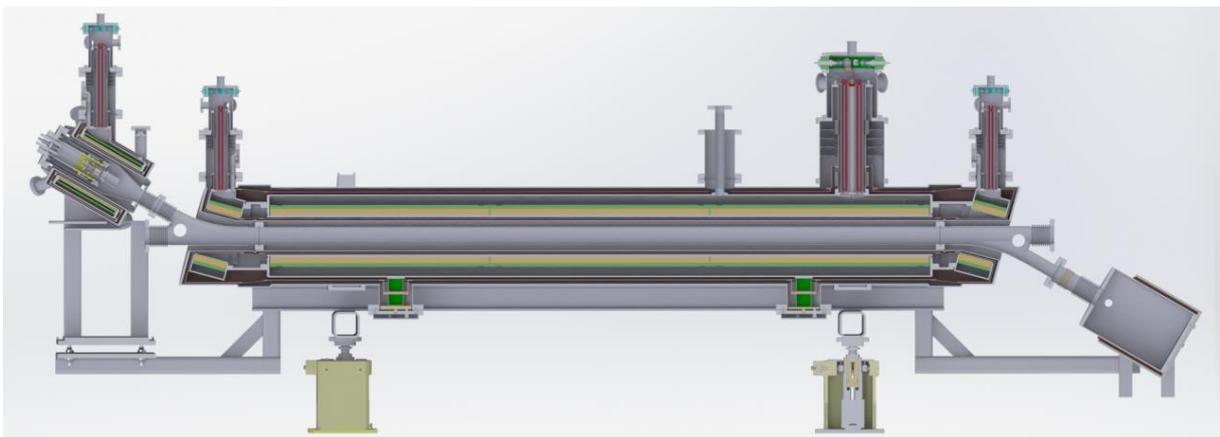

*Figure 17: longitudinal cross section of the HEL. On the left the e-gun cryostat, in the centre the main solenoid and on the right the collector.*

For simplicity all the solenoids and all the correctors are wound using the same Nb-Ti wire. The insulated wire has a rectangular cross section of 0.8 mm x 1.25 mm with a copper to superconductor area ratio of about 4. The critical current measured at 5 T and 4.2 K is at least 400 A. These values are indicative and typical of a wire for MRI applications. They can slightly change in function of the producer that will be chosen.

## 3.2 The central region

The design of the main solenoid is the compromise between two opposite needs. To limit the stored magnetic energy and to reduce the length of the cable and the manufacturing time the solenoid inner diameter should be small. On the other hand the assembly operations of the inner vacuum chamber with the related instrumentation requires space to reach with hands and tools the connecting flanges. The result is a solenoid inner diameter equal to 200 mm.

The main solenoid is 3-m long, it is made in three parts, each of them approximately one metre long connected in series. For each part, the superconducting wire is wound around a cylinder made of fiberglass epoxy (NEMA G11). During the construction a layer of pre-impregnated fibre glass tape is wound in between each layer of superconductor. At the end of the winding the coil is heated in an oven, the tape polymerize and glue together all the components. At the end of this process the three parts are rigid enough to be handled and connected together. Corrector pancake coils are positioned around the main solenoid to steer the electron beam.

The table below gives the main parameters of the central main solenoid.

| Total length | 3 m |
|---|---|
| Number of turns (per layer and per m of length) | 800 |
| Number of layers of superconductor | 25 |
| Total thickness of the coil (after curing) | ~ 27 mm |
| Coil inner diameter | 200 mm |
| Nominal field | 4 T |
| Maximum operation field | 6 T |
| Current at 4 T | 160 A |
| Inductance (total for the 3 m) | 47.4 H |

The solenoid and the correctors are inserted into a helium tank made of austenitic stainless steel. This tank contains the liquid helium that cools the coils at their working temperature, 4.2K. The helium tank has a column to house the current leads, the helium supply line and the helium gas recuperation. The current leads are vapour-cooled and the enthalpy of the cold evaporating helium gas is used to decrease the heat flux through them.

The bending solenoids are wound and cured in the same way as the main solenoid. Around them there are as well corrector coils to steer the electron beam trajectory. Each bending solenoid is housed in a separate helium tank with a dedicated column for helium and current supply. In principle the bending solenoid can be powered at a different current level with respect to the main solenoid. This gives more margin to operate the system. Nevertheless in all the simulations the same current is used in all cases for both the main solenoid and the bending solenoids. The table below gives the main parameters of each of the bending solenoids.

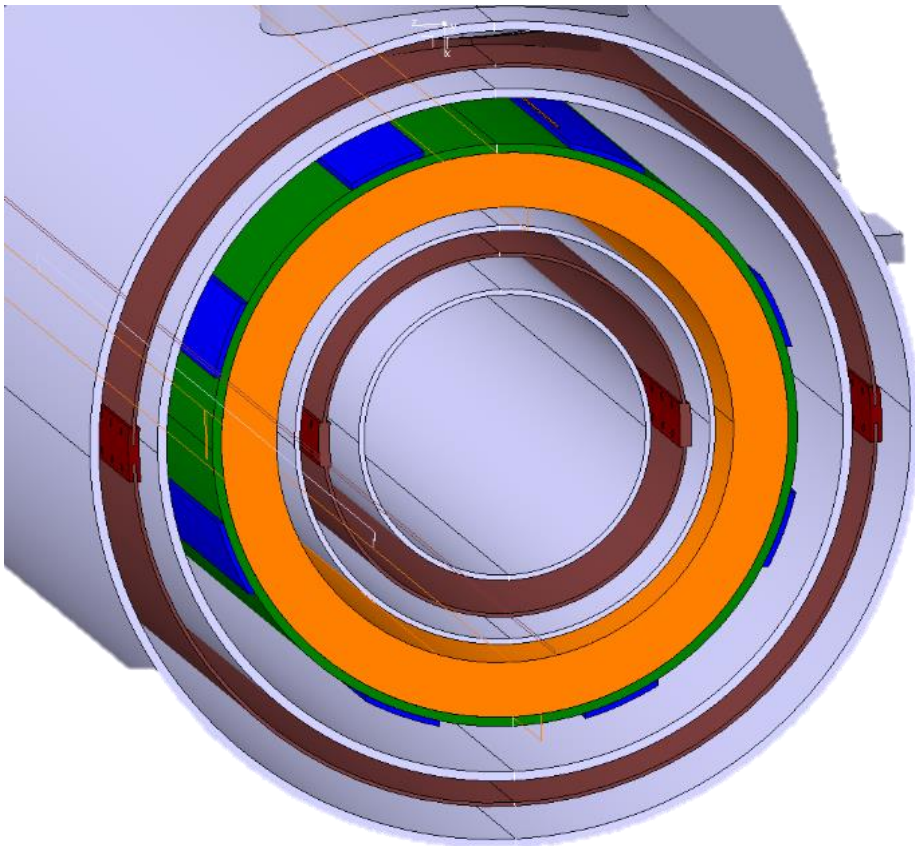

*Figure 18: cross section of the HEL cryostat. In the centre there is the bore for the room temperature vacuum chamber (not shown here). The bore diameter is 135 mm. Then from inside to outside in grey the room temperature vacuum tank, in brown the thermal screen, in grey the helium tank wall, in orange the main solenoid, in blue the corrector coils generating a horizontal and a vertical dipole. Then follow the helium tank wall, the thermal screen and the vacuum tank. The overall external diameter is 454 mm*

| Total length | 0.15 m |
|---|---|
| Number of turns (per layer) | 120 |
| Number of layers of superconductor | 25 |
| Total thickness of the coil (after curing) | ~ 27 mm |
| Coil inner diameter | 220 mm |
| Current (with main solenoid at 4 T) | 160 A |
| Inductance | 2.8 H |

During the powering of the magnetic system the bending coils are attracted by the main solenoid. The force at 4 T is of the order of 45000 N. Separate cryostats for the main and bending solenoids would have required strong supports to keep the coils in position. But strong supports have large cross sections, so they originate heavy heat losses between the fixation points at room temperature and the coils at 4.2 K. On the other hand all the coils cannot be inside the same cryostat otherwise it is not possible to have the physical space to assembly the vacuum chamber inside the system. The chosen solution foresees that the coils are inside separate helium tanks. The tanks are connected by strong plates at 4.2 K that are retaining the magnetic forces transforming them into internal compression forces. The plates are not connected with the outside room temperature structure so that there are no thermal losses. In this way the supports connecting the helium tanks to the outside world are loaded by the structure weight only and their conduction surface can be small. Access to the central beam pipe is possible during the assembly when the connection plates are not yet in place and the vacuum tank is not completely welded.

A copper thermal screen is positioned around the helium tanks. The thermal screen is cooled using helium gas as for the LHC main dipoles.

The supports are very similar to the ones used for the LHC main dipoles. Each support is made of two fiberglass epoxy cylinders in series. The first cylinder is connected on one side to the vacuum tank at room temperature, the second cylinder is connected on the other side to the helium tank. In between the two cylinders there is a plate connected to the thermal screen.

## 3.3 The e-gun region

The electron gun is located inside a dedicated solenoid split in two parts (two separate coils positioned one after the other). The electrons leaving the cathode will first see a magnetic field that can be tuned from 0 T to 0.4 T (the nominal value is 0.2 T). Then they will enter in a zone where the field is determined by a fix current of 94 A in a second solenoid. The fix part define together with the bending solenoids and the main solenoid a stable constant path for the electrons. The tuning part at the beginning of the electron trajectories allow to tune the electron beam diameter.

Characteristics of the fix part of the e-gun solenoid:

| Total length | 0.2 m |
|---|---|
| Number of turns (per layer) | 160 |
| Number of layers of superconductor | 6 |
| Total thickness of the coil (after curing) | ~ 6.5 mm |
| Coil inner diameter | 220 mm |
| Field with 94 A | 0.4 T |
| Maximum operation field | 0.6 T |
| Nominal current | 94 A |
| Inductance | 0.22 H |

Characteristics of the tuneable part of the e-gun solenoid.

| Total length | 0.2 m |
|---|---|
| Number of turns (per layer) | 160 |
| Number of layers of superconductor | 6 |
| Total thickness of the coil (after curing) | ~ 6.5 mm |
| Coil inner diameter | 220 mm |
| Nominal field | 0.2 T |
| Maximum operation field | 0.6 T |
| Current at 0.2 T | 25.9 A |
| Inductance | 0.22 H |

Correction pancake coils are assembled around the 0.4 T solenoid to steer the electron trajectories.

The coils are assembled inside a cryostat with an internal helium tank, surrounded by a thermal screen and the external vacuum tank. The helium feeding and the current lead systems are very similar to the one of the central region cryostat.

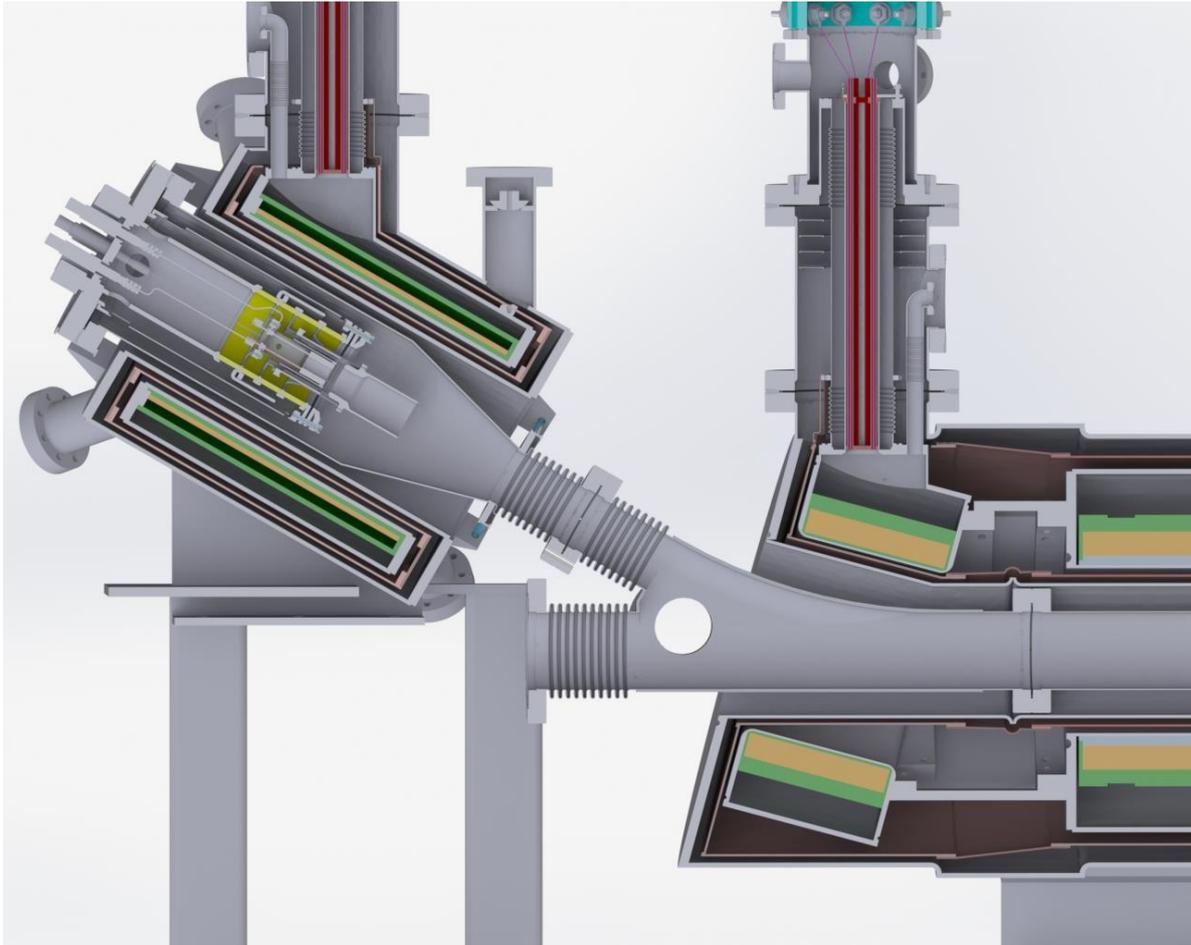

*Figure 19: Longitudinal cross section of the HEL e-gun side. The e-gun is positioned inside the separate cryostat housing the e-gun solenoids (0.4 T and 0.2 T nominal fields). On the left side of the picture there are the intermediate bending solenoid and one part of the main solenoid (in orange).*

## 3.4 The collector region

In case of continuous mode use of the HEL the maximum power deposited on the collector is 50 kW.

*A scheme of the collector shape is shown in*

*Figure 20. We assume that all the surfaces are water cooled. Conservatively, for the thermal assessment, the electrons deposit their energy only on the blue bottom part of 0.2 m of*

Figure 20.

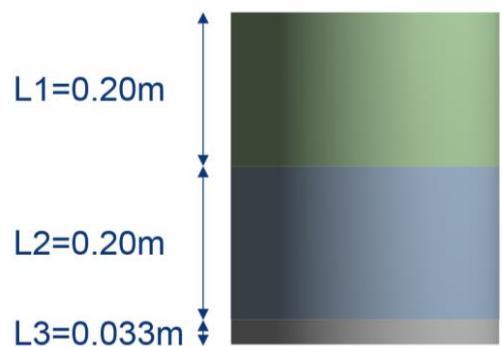

*Figure 20: concept geometry of the collector. All the surfaces are assumed to be cooled.. The electrons deposit their energy on the blue part*

To verify the feasibility of the collector, a thermal model based on the finite difference approach has been written with the code Octave. The basic idea is that the water flows through 8 channels parallel to the axis of the collector. The temperature profiles are shown in Figure 21. The peak temperature is about 84 °C.

The water inlet is at 22 °C with a speed of 1 m/s to avoid corrosion of the copper. The resulting flux is about 8 l/s. The change of water temperature is only ~2 °C. This means that a lower flux is still acceptable.

It takes about 20 min for the collector to reach a steady state. In case of accidental stop of the water flow, the temperature rises with a speed of about 10 °C/min.

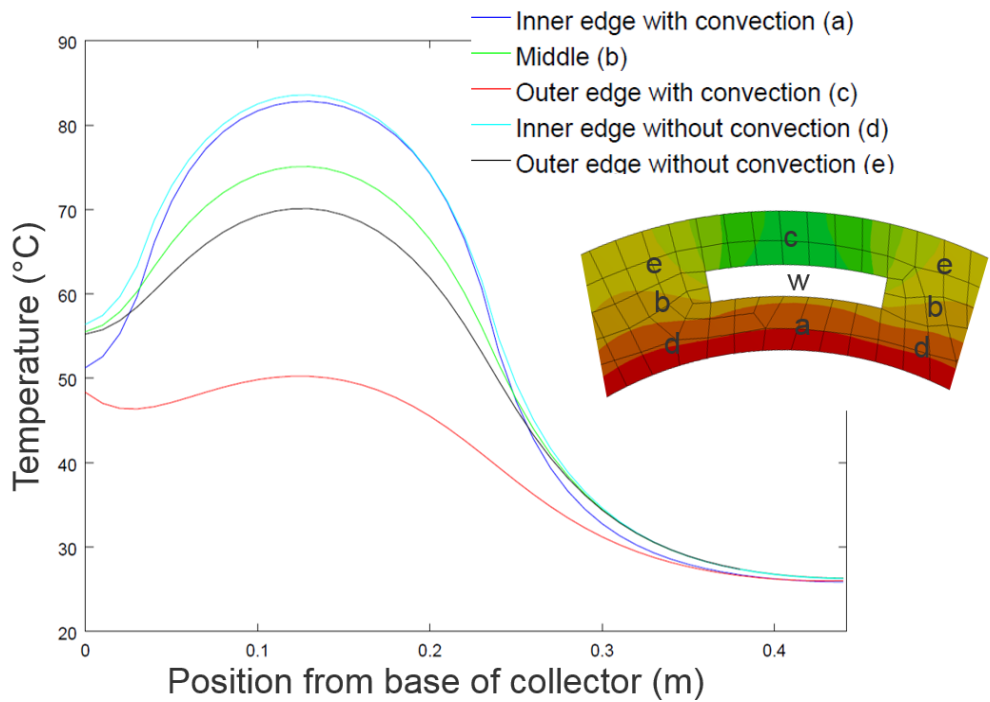

*Figure 21: temperature profiles in some specific positions. The channels are positioned around the collector in a regular pattern.*

The following table gives the main characteristics of the collector:

| Material | ETP Copper |
|---|---|
| Maximum power absorbed | 50 kW |
| Maximum temperature | 85°C |
| Maximum speed of the cooling water | 1 m/s |
| Water flux | 8 l/s |
| Inner diameter | 145 mm |
| External diameter | 175 mm |
| Height | 400 mm |

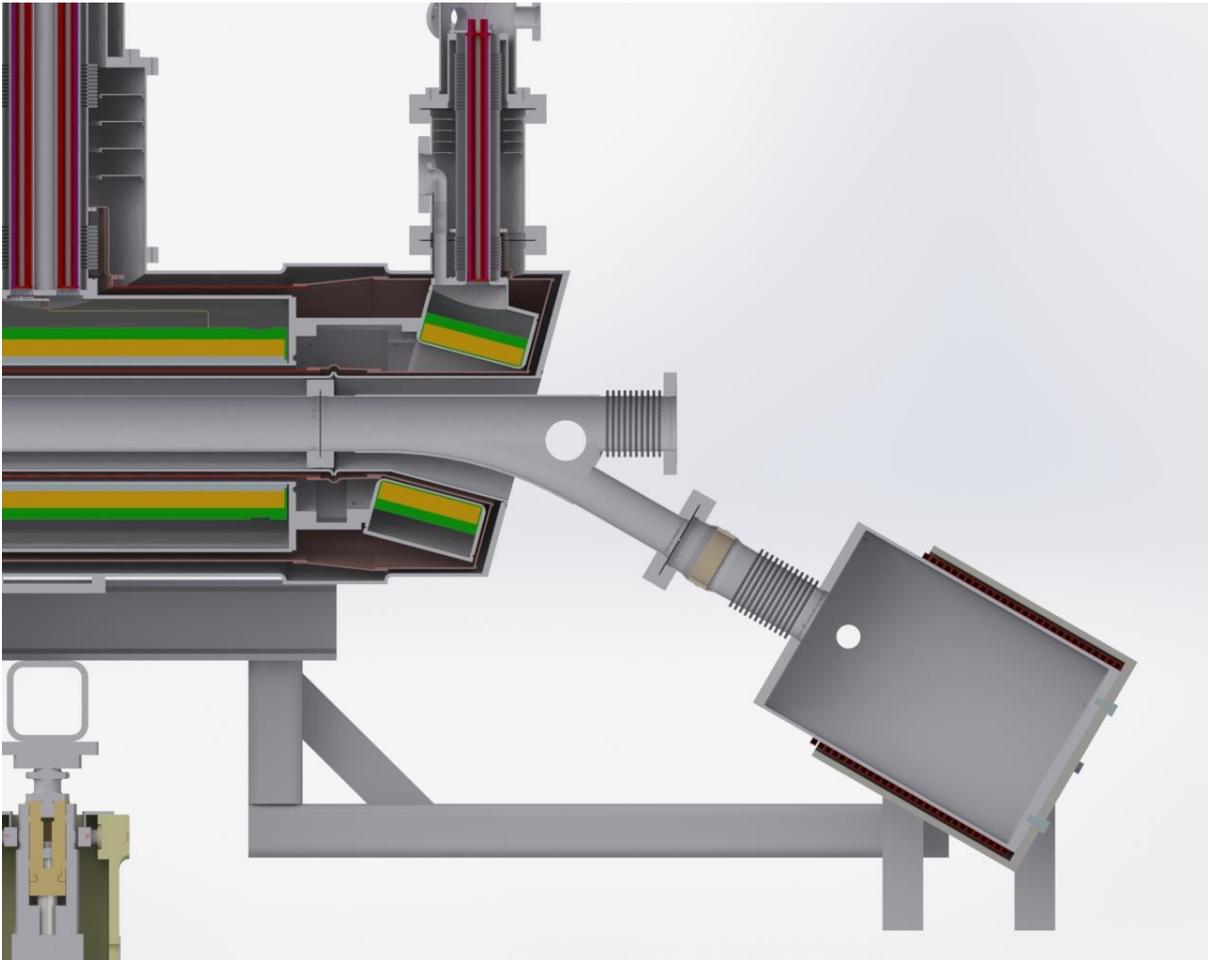

*Figure 22: Longitudinal cross section of the HEL collector side (the collector position and geometry are here not optimized).*

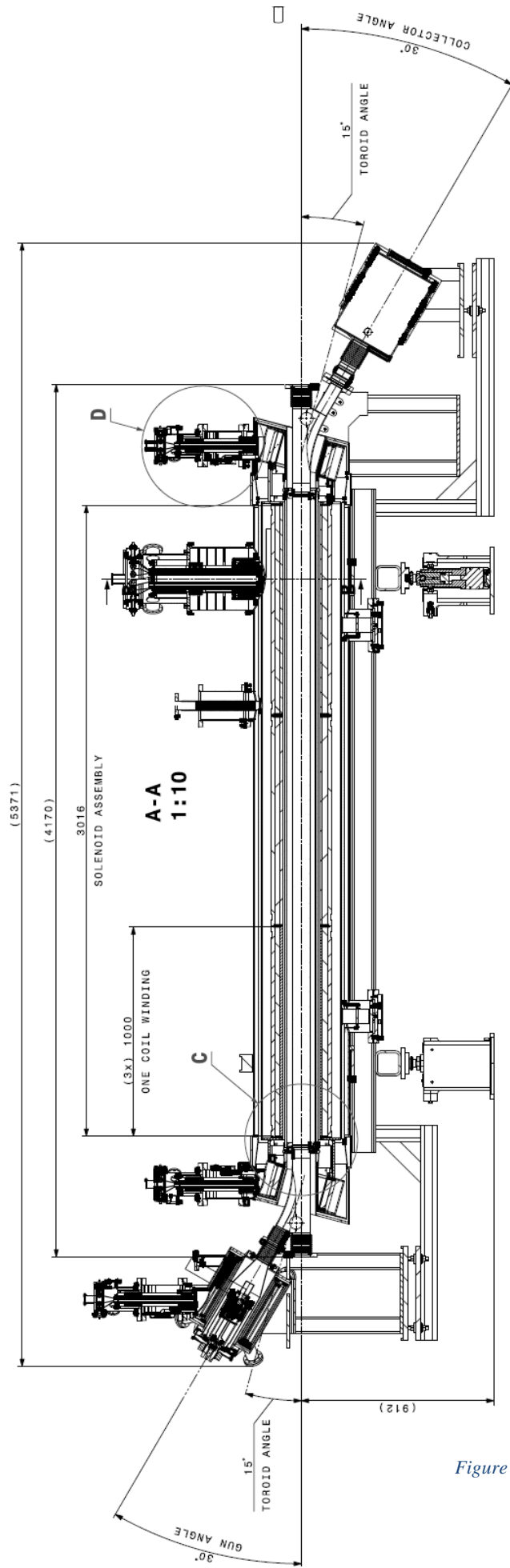

*Figure 23: the HEL main dimensions*

# 4 Conclusions

We presented the work carried out to define a preliminary layout of the Hollow Electron Lenses for HL-LHC. First, we estimated the optimal path of the electron beam and then we designed the magnetic system to keep the particles on their trajectories. The simulations were carried out considering the trajectory of a single electron that represents the centre of the hollow electron beam. This approach is precise enough to allow the complete pre-design of the magnetic system and the associated cryostats and services. The feasibility of a small collector capable of dissipating 50 kW is also demonstrated. The final parameters before preparing the construction drawings shall be defined by a dedicated beam-dynamics computation campaign.

# 5 Acknowledgments

We want to thank for the fruitful discussions and help Giulio Stancari and Miriam Fitterer from Fermilab, the CERN colleagues Stefano Redaelli, Adriana Rossi, Giorgia Gobbi and Antti Kolhemainen,
The mechanical design work has been carried out within a collaboration agreement with the Lapland University of Applied Sciences (LUAS kemi). A great thank to prof. A. Pikkarainen and all the students involved in the project: S. Riekki, J. Harjuniemi, V. Nikkanen, J. Uurasmaa and J. Markkanen. A great thank also to D. Buglass (Oxford University) who carried out the assessment of the collector.